\documentclass[aps,prb,showpacs, twocolumn, superscriptaddress]{revtex4}
\usepackage[utf8]{inputenc}
\usepackage[english]{babel}
\usepackage{graphicx}
\usepackage{color}
\usepackage{bm}
\usepackage{amssymb}
\usepackage[intlimits]{amsmath}
\usepackage{amsmath}

\begin{document}


\title{Properties of the spin liquid phase in the vicinity of the N\'eel - Spin-Spiral Lifshitz transition in frustrated magnets}

\author{Yaroslav A. Kharkov}
\author{Jaan Oitmaa}
\author{Oleg P. Sushkov}
\affiliation{School of Physics, University of New South Wales, Sydney 2052, Australia}

\begin{abstract}
Three decades ago Ioffe and Larkin pointed out a generic mechanism for the formation of a gapped
spin liquid \cite{Ioffe88}. In the case when a classical two-dimensional (2D) frustrated 
Heisenberg magnet
undergoes a Lifshitz transition between a collinear N\'eel phase and a spin spiral phase,
quantum effects usually lead to the development of a spin-liquid phase 
sandwiched between the N\'eel and spin spiral phases.
In the present work, using field theory techniques, we study properties of this universal spin liquid phase.
We examine the phase diagram near the Lifshitz point and calculate the positions of critical points, excitation spectra, and spin-spin
correlations functions.
We argue that the spin liquid in the vicinity of 2D Lifshitz point (LP) is similar to the gapped Haldane phase in integer-spin 1D chains.
We also consider a specific example of a frustrated system with the spiral-N\'eel LP, the $J_1-J_3$ antiferromagnet on the square lattice that manifests the spin liquid behavior.
We present numerical series expansion calculations for this model and compare results of the calculations
with predictions of the developed field theory.
\end{abstract}
\maketitle

\section{Introduction}

Quantum spin liquids (SL) are ``quantum disordered'' ground states of spin systems,
in which zero-point fluctuations are so strong that they prevent conventional magnetic long-range order. 
The main avenues towards realizing SL phases in magnetic systems are frustration and quantum phase 
transitions. \cite{Savary17}
A particularly interesting example of SL is realized by tuning a frustrated magnetic 
system close to a Lifshitz point (LP) that separates collinear and spiral states. 
In the vicinity of the  Lifshitz transition the quantum fluctuations are strongly enhanced, resulting in a plethora of novel intermediate quantum phases \cite{Balents16}. 

A general argument in favour of a universal gapped SL phase near LP  in two-dimensional 
frustrated  Heisenberg antiferromagnets (AF) was first proposed by Ioffe and Larkin \cite{Ioffe88}. 
They showed that in the proximity of the LP  quantum fluctuations destroy long-range spin correlations 
and create a region in the phase diagram with a finite magnetic correlation length.
Subsequent studies found evidence for SL phases in various two-dimensional systems near the LP, 
including Heisenberg models on square and honeycomb lattices with second and third nearest 
neighbor antiferromagnetic couplings \cite{Ferrer93, Capriotti04, Reuther11a,Reuther11b,Zhu13,Zhang13,Bishop15,Oitmaa16,Merino18}.
However, the universality of the SL 
phase near LP, its ubiquitous properties, and the relation of the general argument to
specific Heisenberg models has not previously been addressed.

In the present paper we revisit the Ioffe-Larkin scenario and consider a  
field theory for a quantum Lifshitz transition between collinear and spiral phases in $D=2+1$. 
Disregarding microscopic properties of specific lattice models 
 we focus on the generic infrared physics 
at the LP. 
We develop a field-theoretic description of the $O(3)$ Lifshitz point based on the extended nonlinear 
sigma model.
The nonlinear sigma model provides a unifying theoretical framework that allows us to analyze the phase diagram, calculate  positions of critical points,
excitation spectra, and static spin-spin correlations functions.
We demonstrate  universal scalings of observables  (gaps, position of critical points, etc) in terms of 
the dimensionless SL gap at the LP, $\delta_0$, and show that 
the correlation length in the SL phase scales as $\xi\sim 1/\sqrt{\delta_0}$.
We also argue that the LP spin liquid has a similarity to the gapped Haldane phase\cite{Haldane83} in integer-spin 1D chains.
However, for the 2D SL there is no significant difference between the integer and half-integer spin cases.

A particular example of a system that has a N\'eel-spiral LP and hence manifests the spin liquid behavior is the frustrated antiferromagnetic $J_1-J_2-J_3$ Heisenberg model on the square lattice with the second and third nearest neighbour couplings as well as it's simplified version, the $J_1-J_3$ model.
We perform numerical series expansion calculations for the $J_1-J_3$ model and compare results of the calculations
with predictions of the developed field theory.


The structure of the paper is as follows. In Sec. \ref{sec:field_theory} we introduce 
the effective field theory describing the N\'eel to Spin Spiral Lifshitz point. 
Section \ref{sec:QLP} addresses the quantum LP, quantum fluctuations, and the criterion for quantum `melting'.
Next,  in Sec. \ref{sec:crit_rho_delta} we calculate the spin-wave gap and positions of critical points.
Section \ref{sec:corr} addresses the static spin-spin correlator in the spin liquid phase.
In Sec. \ref{sec:J1-J3} we describe our numerical series calculations for the $J_1-J_3$ model with spin $S=1/2$ and $S=1$
and compare results of these calculations with predictions of the field theory.
Finally our conclusions are presented in Sec. \ref{sec:concl}.

\section{Effective field theory}\label{sec:field_theory}

We start with the following $O(3)$ symmetric Lagrangian describing a transition from the N\'eel to a 
spiral phase in two dimensional antiferromagnets:
\begin{equation}
\mathcal{L} = \frac{\chi_\perp}{2}(\partial_t n_\mu)^2 - \frac{1}{2}n_\mu K(\partial_i) n_\mu,  \quad (n_\mu)^2=1.\label{eq:L}
\end{equation}
Here $\chi_{\perp}$ is the transverse magnetic susceptibility,
$n_{\mu}$ is a unit length vector with $N=3$ components corresponding to the staggered magnetization, $\partial_i$ are the spatial gradients.
The general form of the ``elastic energy'' operator $K(\partial_i)$  in inversion symmetric systems reads 
\begin{equation}\label{eq:K(q)}
K(\partial_i) = -\rho(\partial_i)^2 + \frac{b_1}{2}(\partial_x^4+\partial_y^4)+b_2\partial_x^2 \partial_y^2 + \mathcal{O}(\partial_i^6),
\end{equation}
where we assume that the $n$-field is sufficiently smooth.
 The spin stiffness $\rho$ is the tuning parameter that drives the system across the Lifshitz
transition. 
The spin stiffness is positive in the N\'eel phase, negative  in the spiral phase  and vanishes  at 
the Lifshitz point. The $b$-terms containing higher order spatial derivatives are necessary for stabilization of spiral order at negative $\rho$, and we will assume that $b_{1,2}>0$.
While the kinematic form of the Lagrangian (\ref{eq:L}) is dictated by global symmetries of the system,
 a formal derivation
starting from a frustrated Heisenberg model can be found e.g. in Ref. \cite{Ioffe88} 
Note that in Lagrangian (\ref{eq:L}) we do not take into account topological terms. We will 
discuss their possible role later in the text.

The Lagrangian (\ref{eq:L}) is relevant to a number of models and systems mentioned in the 
Introduction.
Here we would like to mention another example motivated by rare-earth manganite materials 
(Tb,La,Dy)MnO$_3$ (see Ref. \cite{Milstein15}).
These materials have a layered structure with the individual ferromagnetic layers coupled  antiferromagnetically. Due to the antiferromagnetic interlayer coupling the dynamics of the system is described by the second-order time derivative as in usual antiferromagnets in agreement with Eq. (\ref{eq:L}).
 Within each plane there are ferromagnetic nearest neighbour and antiferromagnetic second nearest neighbour Heisenberg interactions leading to an inplane frustration. 
These compounds could be tuned to the N\'eel-Spin-Spiral LP by performing chemical substitution.
Of course real materials are three-dimensional and contain many planes, however thin films can manifest some physics considered here.

\begin{figure}
\includegraphics[scale=0.3]{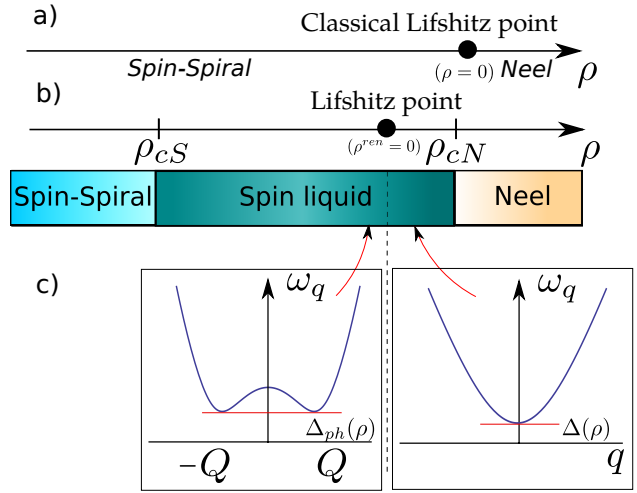}
\caption{ Schematic phase diagram in the vicinity of the Lifshitz transition between collinear antiferromagnetic and spiral states: a) classical Lifshitz transition, b) quantum phase diagram; strong quantum fluctuations in the vicinity of the Lifshitz point result in the intermediate spin liquid phase. c) Excitation energy $\omega_q$ in the spin liquid phase below and above LP.}\label{fig:phase_diagr}
\end{figure}

In the AF phase of (\ref{eq:L}), $\rho > 0$,
the rotational symmetry is spontaneously broken and the N\'eel vector has a nonzero expectation value, 
e.g. is directed along the $z$ axis $\langle \bm n \rangle=\bm e_z$.
In the spin spiral phase, with $\rho<0$, there is an incommensurate ordering 
\begin{equation}\label{eq:n}
\bm n(\bm r) = \bm {e_1} \cos(\bm{Qr}) + \bm {e_2} \sin(\bm{Qr}),
\end{equation}
where $\bm{e_{1,2}}$ are orthogonal unit vectors and $\bm Q$ is the pitch of 
the spiral. 
For $b_1\leq b_2$ the spiral wave vector is directed along $x$ or $y$: 
$\bm Q= (\pm Q,0), (0,\pm Q)$, where $Q^2=|\rho|/b_2$. In the opposite case $b_1>b_2$ the wave vector is directed along the main diagonals: $\bm Q= \frac{1}{\sqrt{2}}(\pm Q,\pm Q), \frac{1}{\sqrt{2}}(\pm Q,\mp Q)$, where $Q^2 = 2|\rho|/(b_1+b_2)$.
The relation between the coefficients $b_1$ and $b_2$ depends on the specific choice of the lattice model. 
In the ``isotropic'' case, $b_1=b_2$, the system has additional rotational degeneracy  in the momentum space due to the arbitrary orientation of wave vector $\bm Q$.
The additional degeneracy can destabilize spiral states and result in quantum spin liquid states that have been predicted for 3D antiferromagnets.\cite{Bergman07}
In the present paper we will stay away from this special critical point.
The classical phase diagram is shown schematically in Fig.\ref{fig:phase_diagr}a.

We would like to make a comment regarding Lagrangian (\ref{eq:L}).
Parameters of any field theory depend on the momentum and energy scales that is described by renormalization group procedure.
We assume that parameters in (\ref{eq:L}),(\ref{eq:K(q)})
 are fixed at the ultraviolet cutoff
$\Lambda \approx 1$, where unity corresponds to the inverse lattice spacing.
Quantum fluctuations at scales larger than $\Lambda$ but smaller
than the boundary  of  magnetic  Brillouin  zone lead to a renormalization of the parameters $\rho\rightarrow\rho^{ren}$, $ b_{1,2}\rightarrow b_{1,2}^{ren}$, $\ldots$. Therefore, the values of the parameters in
(\ref{eq:L}),(\ref{eq:K(q)})
can be different from those naively derived using spin wave theory. 
As was pointed out by Ioffe and Larkin \cite{Ioffe88} this renormalization is especially relevant for the
spin stiffness. The correction to the spin stiffness arises due to the the $b$-terms in (\ref{eq:K(q)}).
The easiest way to understand the correction\cite{Polyakov75}  is to consider the N\'eel phase
and decompose the order parameter into two transverse components and a longitudinal component

\begin{equation}
\bm n = (\bm \pi,n_z), \quad n_z=\sqrt{1-\bm \pi^2}\approx 1- \bm \pi^2/2 \ . 
\end{equation}
Hence the following contribution from the $b$-term arises
\begin{eqnarray}
\label{nz}
\partial^2 n_z  \partial^2 n_z
\sim b \ (\partial^2 \bm{\pi}^2 ) ( \partial^2 \bm\pi^2 )\ .
\end{eqnarray}
The field  $\bm\pi$ has fluctuations with momenta smaller than $\Lambda$, ${\bm \pi}_<$,
and fluctuations with momenta larger  than $\Lambda$, ${\bm \pi}_>$, ${\bm \pi}={\bm \pi}_<+{\bm \pi}_>$.
Substitution in (\ref{nz}) and averaging over high energy fluctuations gives
\begin{eqnarray}
\label{nz1}
b \ (\partial^2 \bm{\pi}^2 )( \partial^2 \bm\pi^2 )
\to
b \ (\partial \bm{\pi}_<)^2  \langle(\partial \bm\pi_>)^2\rangle 
=\delta\rho_{\Lambda}(\partial \bm{\pi}_<)^2\ .
\end{eqnarray}
Note, when averaging $(\partial^2 \bm{\pi}^2 )\times (\partial^2 \bm\pi^2)$ each multiplier must
contain the high ($\bm\pi_>$) and the low ($\bm\pi_<$) energy components. The terms with one multiplier containing
only the high energy and another only the low energy components give rise to a total derivative contributions to the Lagrangian and can be neglected.
Equation (\ref{nz1}) demonstrates a positive correction to the spin stiffness.
Therefore quantum fluctuations always extend the N\'eel phase
compared to the prediction of  spin-wave theory that is indicated in Panel b of Fig. \ref{fig:phase_diagr}.
The Lifshitz point in the quantum case is shifted to the left compared to the Lifshitz
point in the classical case.
In the quantum case the Lifshitz point is ``buried'' in the spin liquid phase.
Nevertheless, it is unambiguously defined as we discuss in the following Sections.

\section{Quantum Lifshitz point: the phase diagram and the spin liquid gap}\label{sec:QLP}
Quantum fluctuations destroy the  classical N\'eel to spin-spiral Lifshitz transition \cite{Ioffe88}.
Let us calculate  local staggered magnetization $n_z$ when approaching the LP from 
the N\'eel phase.
Representing the staggered magnetization  as
$\langle n_z \rangle \approx 1 - \frac{1}{2}\langle \bm \pi^2 \rangle$, we obtain

\begin{eqnarray}\label{eq:pi2}
\langle \bm \pi^2  \rangle \approx  (N-1) \sum_q \int \frac{id\omega }{(2\pi)} \frac{1}{\chi_\perp \omega^2 - K(\bm q)  + i0}  \nonumber \\ = 
(N-1) \int \frac{d^2q}{(2\pi)^2}  \frac{1/\chi_\perp}{2\omega_q}, 
\end{eqnarray}
where $\omega_q = \chi_\perp^{-1/2}\sqrt{\rho q^2 + b_1/2 (q_x^4+q_y^4) + b_2q_x^2q_y^2}$.
In the vicinity of the LP, $\rho\rightarrow 0$, the integral (\ref{eq:pi2}) is logarithmically divergent, $\langle \bm \pi^2\rangle \propto \ln \left(\frac{\Lambda}{\sqrt\rho}\right)$, where $\Lambda$ is the ultraviolet momentum  cutoff.  
Hence at some critical value of the spin stiffness $\rho=\rho_{cN}$ the staggered 
magnetization $\langle n_z \rangle$ vanishes, indicating a transition to the spin liquid phase. 
In the spin liquid phase, $\rho<\rho_{cN}$, a gap $\Delta$ must open to regularize 
the integral in Eq. (\ref{eq:pi2}) 
\begin{eqnarray}\label{eq:omega_q}
&&\omega_q\rightarrow \sqrt{\omega_q^2+\Delta^2} \nonumber\\ 
&&=\sqrt{\Delta^2 + \chi_\perp^{-1}[\rho q^2 + b_1/2 (q_x^4+q_y^4) + b_2q_x^2q_y^2]}.
\end{eqnarray} 

Opening of the gap indicates an existence of a spin liquid phase at which the long range AF order is lost and the order parameter correlations are exponentially decaying.
Importantly, this is a generic gapped spin liquid originating from long range fluctuations and is unrelated to a spin-dimer ordering.
The SL gap is zero, $\Delta=0$, at the critical point $\rho_{cN}$ and the gap increases when we proceed deeper into the spin liquid phase. 
The SL phase stretches across a finite window $[\rho_{cS},\rho_{cN}]$ in the vicinity of the LP, as depicted in Fig. \ref{fig:phase_diagr}b.

The elementary spin excitations in the AF phase are two gapless Goldstone modes - transverse spin-waves and a 
massive longitudinal ('Higgs') mode. Due to the unit length constraint ($\bm n^2=1$) the Higgs mode has a very 
large energy and can be disregarded.
In the spiral phase there are three Goldstone modes: a sliding mode and two out of plane excitations. These three 
modes correspond to the three Euler angles defining the orientation of the $(\bm e_1, \bm e_2, \bm e_3 )$ triad, 
where $\bm e_3 = [\bm e_1 \times \bm e_2]$.\cite{Azaria90, Milstein15}.

The excitation modes (\ref{eq:omega_q}) in the SL phase  are three-fold degenerate due to $O(3)$ rotational invariance of the model. Above the LP ($\rho>0$) the minimum of dispersion is located at $q=0$, whereas below the LP ($\rho<0$) the dispersion has four degenerate minima at the 'spiral' wave vectors  $q=\bm Q$.
The evolution of the dispersion across the LP is schematically shown in Fig. \ref{fig:phase_diagr}c.
The change of the shape of the dispersion indicates the Lifshitz point.

The location of this critical point $\rho_{cN}$  can be found by imposing the condition $\langle n_z\rangle \rightarrow 0$, 
which naively provides the following criterion for the transverse spin fluctuations $\langle \bm\pi^2  \rangle_c \approx 2 $. 
This critical value for $\langle \bm\pi^2 \rangle$ is largely overestimated and it is not consistent with the 
unit length constraint.
One can find a more accurate value of $\langle \bm\pi^2 \rangle_c$ by accounting for the next order terms in the Taylor 
series expansion of $n_z = \sqrt{1-\bm\pi^2}$  (see Appendix \ref{sec:nz_Taylor}), or alternatively by using the $1/N$ expansion for $O(N)$ theory. 
The $1/N$ expansion has been extensively applied to describe quantum antiferromagnets. For the most relevant examples see Refs. \cite{Read90, Chubukov94, Affleck89}.
In the $1/N$ expansion approach we lift the hard constraint $\bm n^2=1$ by introducing a Lagrange multiplier
\begin{equation}\label{eq:L+lmbda}
\mathcal{L} \rightarrow \mathcal{L} - \lambda (\bm n^2 - 1).
\end{equation}

After integrating out the $\bm n$ field in the new Lagrangian (\ref{eq:L+lmbda}), we obtain an effective Lagrangian depending only on the auxiliary field $\lambda$:
\begin{equation}\label{eq:L_eff}
\mathcal{ L}_\lambda = N tr \ln (-\chi_\perp\partial_{tt} - K(\bm q) - \lambda) + \lambda.
\end{equation}
We can find the saddle point in the Lagrangian $\mathcal L_\lambda$ by calculating the variational derivative in  (\ref{eq:L_eff}) with respect to $\lambda$ and regarding $\lambda$ as a constant, $\lambda=\chi_\perp\Delta^2$: 
\begin{equation}\label{eq:N_gap_cond}
N \sum_q \int \frac{id\omega}{(2\pi)}\frac{1}{\chi_\perp(\omega^2 - \Delta^2) - K(\bm q) } = 1.
\end{equation}

The Lagrange multiplier in Eq. (\ref{eq:N_gap_cond}) has the meaning of the spin gap. Equation (\ref{eq:N_gap_cond}) determines the evolution of the gap $\Delta(\rho)$ with the spin stiffness in the SL phase.
 Comparing Eq. (\ref{eq:N_gap_cond}) with Eq. (\ref{eq:pi2}) we conclude that at the boundary between SL and AF phases 
$ \langle \bm\pi^2\rangle_c = (N-1)/N=2/3$.
This criterion is quite natural for the $O(3)$ symmetric  quantum critical point separating N\'eel and SL states.
Nevertheless, this criterion underestimates $\langle \bm\pi^2\rangle_c$.
One can see this from the example of the $S=1/2$ 2D Heisenberg model on the square lattice.
A textbook expression for the staggered magnetization is well known
\begin{eqnarray}
\label{hm}
\langle n_z\rangle =2\langle S_z\rangle= 1-2\int_{MBZ}\frac{d^2q}{(2\pi)^2}\left(\frac{1}{\sqrt{1-\gamma_q^2}}-1\right) ,
\end{eqnarray}
where $\gamma_q=\frac{1}{2}(\cos q_x+\cos q_y)$, and integration is performed over the magnetic Brillouin zone.
In the limit $q<1$ Eq. (\ref{hm}) is consistent with (\ref{eq:pi2}) since in this case $\chi_{\perp}=1/8J$ and
$\omega_q/J \approx \sqrt{2}q$, where $J$ is the Heisenberg AF coupling. Integration over $q$ in (\ref{hm}) gives a well known result $\langle n_z\rangle \approx 2\times0.305$
which corresponds to $\langle \bm\pi^2\rangle \approx 0.78$ in the equation
$\langle n_z \rangle \approx 1 - \frac{1}{2}\langle \bm \pi^2 \rangle$.
The integration in the corresponding long-wavelength approximation (\ref{eq:pi2}) with $N=3$,
$\chi_{\perp}=1/8J$, $\omega_q\approx \sqrt{2}J q$ and the ultraviolet cutoff $\Lambda=1$
gives a close value $\langle \bm\pi^2\rangle \approx 0.89$.
Both values are above 2/3 and we know that the long range AF order in the unfrustrated Heisenberg model
still persists.
Based on this analysis we estimate the critical value of fluctuation as
\begin{eqnarray}
\label{pc}
\langle \bm\pi^2\rangle_c \approx 1. 
\end{eqnarray}

Equation (\ref{pc}) is an analogue of the Lindemann criterion for quantum melting of long range magnetic order
in 2D quantum magnets.
Our approach implicitly violates rotational invariance, but it allows us to calculate approximately
the positions of critical points and the value of the spin liquid gap.

The spin liquid gap $\Delta$ is determined by Eqs. (\ref{eq:pi2}) and (\ref{eq:omega_q})
from the condition $\langle \bm\pi^2\rangle = \langle \bm\pi^2\rangle_c \approx 1$.
At $\rho> 0$ (the N\'eel side of LP) $\Delta$ coincides with the physical gap.
On the spiral side of LP, $\rho<0$, the physical gap corresponds to the excitation energy at the ``spiral'' 
wave vector $\bm Q$: $\Delta_{ph} = \min \omega_q = \sqrt{\Delta^2+\frac{1}{\chi_\perp}K(\bm Q)}$,
see Fig. \ref{fig:phase_diagr}c.
This gap is closed at the spin-spiral-SL critical point. Therefore, the position of this critical point $\rho_{cS}$ is determined from the following
 two equations 
\begin{equation}
\begin{cases}
 2 \sum_{q< \Lambda} \int \frac{id\omega }{(2\pi)} \frac{1}{\chi_\perp(\omega^2 - \Delta^2) - K(\bm q) + i0}=1,\\
\Delta^2_{ph}=\Delta^2 + \frac{1}{\chi_\perp}K(Q) = 0.
\end{cases}
\end{equation}
At $\rho<\rho_{cS}$, the magnon Green's function  acquires a pole at imaginary frequency 
$\omega=\pm i \sqrt{|\Delta^2 + K(Q)/\chi_\perp|}$. This is the indication of an instability of the SL 
phase towards condensation of a static spiral with the wave vector $\bm Q$.

It is instructive to draw an analogy between the SL physics at 2D Lifshitz point and the one-dimensional Haldane spin chain.
A condition similar to (\ref{eq:N_gap_cond}) determines the value  of the Haldane gap. \cite{Affleck89}
Indeed, the integer spin $S$ Heisenberg model in the continuous limit can be mapped to the $O(3)$ 
relativistic nonlinear sigma model in $D=1+1$.\cite{Haldane83} 
The model parameters are the speed of the magnon, $c=\sqrt{\rho/\chi_\perp}=2J S$, and the
 transverse magnetic susceptibility, $\chi_\perp = 1/4J$ ($J$ is the Heisenberg coupling constant).
Proceeding by analogy with (\ref{eq:pi2}) we find the fluctuations of the spin in the Haldane model
\begin{equation}
\label{eq:NLSM_Haldane}
\langle \bm\pi^2\rangle_c = 2\int_0^\Lambda \frac{dq}{2\pi} \frac{1}{2\chi_\perp\sqrt{c^2q^2 
+ \Delta^2}} \approx \frac{1}{2\pi c\chi_{\perp}} \ln \frac{c\Lambda}{\Delta},
\end{equation}

As we already discussed, the ultraviolet cutoff is $\Lambda\approx 1$.
The logarithmically divergent $\langle \bm \pi^2 \rangle$ in the Haldane model is analogous to the 
log-divergence in (\ref{eq:pi2}) at the LP.
Numerical values of the Haldane gaps for  $S=1$ and $S=2$ are known from DMRG calculations: see e.g. 
Ref. \cite{Renard03}, $\Delta_{S=1}/J \approx 0.41$, $\Delta_{S=2}/J \approx 0.08$.
Taking these values of the gap Eq.(\ref{eq:NLSM_Haldane}) we obtain the following critical values of
fluctuations, $\langle \bm\pi^2 \rangle_c \approx 0.5$ (for $S=1$) and $\langle \bm\pi^2 \rangle_c \approx 0.6$ 
(for $S=2$), which
are smaller than (\ref{pc}). We believe that the difference is due to different dimensionality.
While DMRG is more reliable it is interesting to note that the  renormalization group  analysis \cite{Affleck89} 
for the Haldane chain gives $\langle \pi^2\rangle_c = 1$.

The differences in the values of $\langle \pi^2\rangle_c$ is not crucial when making comparisons between 1D and 2D systems.
However, it is well known that properties of the spin chains with half-integer and integer spins are very different. 
The gapped SL phase in 1D appears only in the integer spin chains, while in contrast the excitations of 
half-integer spin chains are gapless spinons in agreement with the Lieb-Shultz-Mattis theorem. \cite{Lieb61}
We believe that the 2D spin liquid in the vicinity of LP point is generic and 
independent of the spin value.
The Lieb-Shultz-Mattis theorem states that in systems with half-integer spin per unit lattice cell and 
full rotational $SU(2)$ symmetry the excitations are gapless or otherwise the ground state of the system 
is degenerate.  The theorem was initially formulated for $D=1+1$ systems and later generalized for higher spatial 
dimensions \cite{Hastings04}.
Technically in $D=1+1$ the dramatic difference between integer and half integer spin is due to the topological Berry 
phase term which is not included in the Lagrangian (\ref{eq:L}).\cite{Haldane83}
Topological effects in $D=2+1$  correspond to skyrmions or merons.\cite{Takayoshi16} 

In principle topological configurations  become more important when approaching the Lifshitz point. \cite{Kharkov17}
However such topological solutions are unstable within the model (\ref{eq:L}). Using scaling arguments one can see that due to the fourth spatial derivative term in the Lagrangian (\ref{eq:L}) the energy of localized skyrmions at LP behaves as $\sim b_{1,2}/R^2$, where $R$ is the skyrmion radius. 
Therefore any localized skyrmions energetically prefer to have large size $R\rightarrow\infty$ and only contribute to the boundary terms.
Although the topological solutions might play a role to reconcile with the Lieb-Shultz-Mattis theorem, these configurations are statistically irrelevant in the bulk.

\section{Positions of N\'eel-Spin liquid and Spin-Spiral-Spin liquid critical points}\label{sec:crit_rho_delta}
In order to make our calculations more specific and having in mind comparison with the $J_1-J_3$ model, in this Section we set $b_2=0$.
It is convenient to introduce dimensionless spin stiffness and dimensionless gap parameters
\begin{eqnarray}
\label{rd}
&&\bar\rho = \frac{2\rho}{b_1},\quad \delta = \sqrt{\frac{2\chi_\perp}{b_1}}\Delta\ .
\end{eqnarray}

At negative $\rho$ the spiral wave vector is directed along the main diagonals 
$\bm Q = \frac{1}{\sqrt{2}}(Q,\pm Q)$,
\begin{eqnarray}
\label{qdd}
Q^2=|\bar\rho| \ .
\end{eqnarray}
As we already discussed in Section \ref{sec:QLP} the condition of criticality reads
\begin{eqnarray}\label{eq:pi2_b2=0}
\langle{\bm\pi^2}\rangle_c\approx 1\approx
\frac{\sqrt{2}}{(4\pi^2)\sqrt{\chi_{\perp}b_1}}
\int\frac{d^2q}{\sqrt{{\bar \rho}q^2+q_x^4+q_y^4+\delta^2}} \ .
\end{eqnarray}

First, we determine the gap  exactly at the LP, $\delta_0=\delta(\rho= 0)$.
For $\delta_0 \ll 1$ the solution of (\ref{eq:pi2_b2=0}) is
\begin{eqnarray}
\label{eq:delta_0}
\delta_0=1.7 \Lambda^2 e^{- \frac{2\sqrt{2}\pi}{\zeta}  \sqrt{\chi_{\perp}b_1} }.
\end{eqnarray}
The constant $\zeta$ in the exponent is given by the angular part of the $q$-integral 
$\zeta = \frac{2}{\pi}K\left(\frac{1}{2}\left[1-\frac{b_2}{b_1}\right]\right)$, 
where $K(m) = \int_0^{\pi/2} d\phi\frac{1}{\sqrt{ 1 - m \sin^2{\phi} }}$ is the complete elliptic integral.
In the specific case under consideration, $b_2 = 0$, $\zeta = \frac{2}{\pi}K(1/2)\approx 1.18$.  
The numerical prefactor $A=1.7$ in (\ref{eq:delta_0}) is found by performing a least-squares 
fitting of  the integral in Eq. (\ref{eq:pi2_b2=0}).
While Eq. (\ref{eq:delta_0}) is derived for $\delta_0 \ll 1$, however direct numerical integration in (\ref{eq:pi2_b2=0}) shows that (\ref{eq:delta_0}) practically works up to 
 $\delta_0 \leq 0.6 - 0.7$.

In order to determine the position of the N\'eel critical point $\rho_{cN}$ we evaluate
 the integral in (\ref{eq:pi2_b2=0}) at $\delta \ll {\bar{\rho}}\ll 1$,
\begin{eqnarray}
\label{qf_main}
\frac{1}{2\pi}
\int\frac{d^2q}{\sqrt{{\bar {\rho}}q^2+q_x^4+q_y^4+\delta^2}}
\approx \frac{\zeta}{2} \ln\left(\frac{2.9 \Lambda^2}{\bar{\rho}}\right)- \frac{\delta}{\bar{\rho}} 
\end{eqnarray}

The condition $\delta=0$ gives the position of the N\'eel-SL critical  point $\bar\rho_{cN}$:
\begin{eqnarray}
\label{fi1_main}
{\bar {\rho}_{cN}}\approx 2.9 \Lambda^2 e^{- \frac{2\sqrt{2}\pi}{\zeta}  \sqrt{\chi_{\perp}b_1} } 
\approx 1.65\delta_0 \ .
\end{eqnarray}
According to (\ref{qf_main}) in the vicinity of the N\'eel-SL critical point, ${\bar {\rho}} < {\bar {\rho}_{cN}}$,
the gap grows linearly as $\delta \approx  0.64({\bar {\rho}_{cN}}-{\bar {\rho}})$, that corresponds to a mean-field prediction. 

The spin stiffness $\rho_{cN}$ at the transition point from the N\'eel phase to the spin liquid phase is small but still finite.
Therefore, we believe that the transition belongs to the standard $O(3)$ universality class, the same as that in the bilayer 
quantum antiferromagnet, see e.g. Ref. \cite{Shevchenko2000}
The correct critical index for $O(3)$ transition is $\nu\approx0.7$, which implies
$\delta \propto ({\bar {\rho}_{cN}}-{\bar {\rho}})^{\nu}$.

On the side of negative spin stiffness, $\bar\rho_{cS}<\bar\rho<0$, the dimensionless physical gap reads
\begin{equation}\label{eq:delta_ph}
 \alpha = \sqrt{\frac{2\chi_\perp}{b_1}}\Delta_{ph} =  \sqrt{\delta^2 -  \bar\rho^2/2}.
\end{equation}

The condition $\alpha=0$ determines the position of the spin-spiral to SL critical point $\rho_{cS}$. 
Calculating the integral in (\ref{eq:pi2_b2=0})
 at $\alpha \ll Q^2 \ll 1$ we find
\begin{eqnarray}
\label{qf7_main}
&&\frac{1}{2\pi}
\int\frac{d^2q}{\sqrt{Q^4/2-Q^2q^2+q_x^4+q_y^4+\alpha^2}}\nonumber\\
&&\approx \zeta \ln\left(\frac{5.4 \Lambda}{Q}\right) -2 \frac{\alpha}{Q^2} \ . 
\end{eqnarray}
The condition $\alpha=0$ gives the position of the critical point $\bar\rho_{cS}$:
\begin{equation}
\label{eq:rho_cS}
\bar\rho_{cS} = - Q^2 \approx - 15\delta_0  \ .
\end{equation}
The gap in the vicinity of this critical point is $\alpha = 0.27 (\bar\rho-\bar\rho_{cS})$.
This is a mean-field result and we believe that the transition at $\rho_{cS}$ does not belong
to a standard universality class.

\begin{figure}
\includegraphics[scale=0.3]{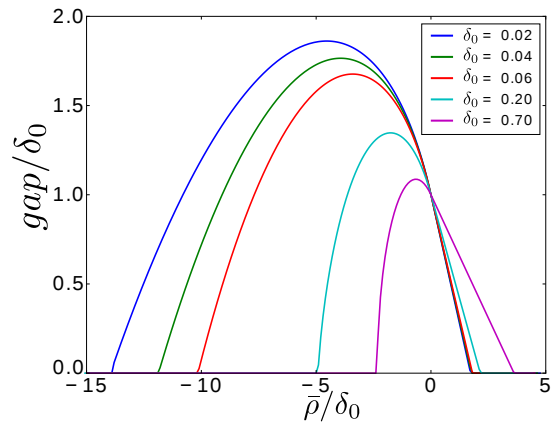}
\caption{ Dimensionless spin liquid gap versus spin stiffness
for different values of $\delta_0$. 
 }
\label{fig:gaps_spins}
\end{figure}

The dimensionless gap found by numerical  solution of Eq. (\ref{eq:pi2_b2=0}) for different values of $\delta_0$
 in the entire SL 
region $\rho_{cS}<\rho<\rho_{cN}$ is presented in Fig. \ref{fig:gaps_spins}.
From this figure we conclude that  asymptotic solutions given by Eqs. (\ref{fi1_main}) 
and (\ref{eq:rho_cS}) become valid only at sufficiently small values of $\delta_0$ 
(i.e large values of S): Eq. (\ref{fi1_main}) is valid at $\delta_0 \lesssim 0.2$ and
Eq. (\ref{eq:rho_cS}) is valid only for very small gaps, $\delta_0 \lesssim 0.02$.
The asymmetry between $\rho_{cS}$ and $\rho_{cN}$ evident from Fig. \ref{fig:gaps_spins}
is due to stronger quantum fluctuations in the spiral ($\rho<0$) region compared to 
the $\rho>0$ domain.

An alternative method to determine $\bar{\rho}_{cS}$ is to approach the spiral-SL critical point from the spiral phase and find the condition when quantum fluctuations melt the spiral.
The fluctuations of spiral consist of the out-of-plane $h({\bm r},t)$ and in-plane modes $\phi(\bm r,t)$, can be parametrized in the form
\begin{eqnarray}
\label{wh0}
{\vec n}=(\sqrt{1-h^2}\cos({\bm Q}\cdot{\bm r}+\phi),\sqrt{1-h^2}\sin({\bm Q}\cdot{\bm r} + \phi),h) \ .
\end{eqnarray}
The total quantum fluctuation orthogonal to the spin alignment in the spiral state reads
\begin{eqnarray}
\label{qf}
&&\langle \bm\pi^2 \rangle = \langle \phi^2 \rangle+\langle h^2 \rangle, \\
&&\langle \phi^2 \rangle=
\frac{1}{(4\pi^2)\sqrt{2\chi_{\perp}b_1}}
\int\frac{d^2q}{\sqrt{2Q^2q^2+q_x^4+q_y^4}},\nonumber\\
&&\langle h^2 \rangle=
\frac{1}{(4\pi^2)\sqrt{2\chi_{\perp}b_1}}
\int\frac{d^2q}{\sqrt{Q^4/2-Q^2q^2+q_x^4+q_y^4}}.\nonumber
\end{eqnarray}

The denominators in the integrals for $\langle\phi^2\rangle$ and $\langle h^2\rangle$ in (\ref{qf}) represent the dispersions for the Nambu-Goldstone excitations: the sliding mode and the out of plane mode, see details in Appendix \ref{sec:append_spiral}.
Evaluating the integrals with logarithmic accuracy, we obtain
\begin{eqnarray}
\label{qf2}
&&\langle \bm\pi^2 \rangle \approx \frac{1}{(2\pi)\sqrt{2}\sqrt{\chi_{\perp}b_1}}
\zeta\ln\left(\frac{6.5 \Lambda^2}{Q^2}\right).
\end{eqnarray}
Now, applying the same criterion for the critical point, $\langle \bm\pi^2 \rangle_c \approx 1$,
we find the critical $\bar\rho_{cS}$
\begin{eqnarray}
\label{qc}
\bar\rho_{cS} \approx -6.5 \Lambda^2 e^{- \frac{2\sqrt{2}\pi}{\zeta}  \sqrt{\chi_{\perp}b_1} } \approx -4\delta_0.
\end{eqnarray}

The prefactor in (\ref{qc}) is significantly smaller then the prefactor in Eq.(\ref{eq:rho_cS}).
This emphasizes the fact that our calculation is only approximate. 
Pragmatically this uncertainty is not very significant.
We already pointed out that Eq. (\ref{eq:rho_cS}) is valid only for extremely small gaps, 
$\delta_0 \lesssim 0.02$. At larger values of $\delta_0$ the position of the critical
point $\rho_{cS}$ is different from (\ref{eq:rho_cS}), see Fig. \ref{fig:gaps_spins}.
Numerical evaluation of (\ref{qf}) combined with the criticality condition (\ref{pc})
gives  the following locations of the critical points $\rho_{cS}$:
${\bar \rho}_{cS}/\delta_0=-3.7$ at $\delta_0=0.06$;
${\bar \rho}_{cS}/\delta_0=-3.6$ at $\delta_0=0.2$;
${\bar \rho}_{cS}/\delta_0=-2.2$ at $\delta_0=0.7$.
Comparing these values with positions of the critical point that
follow from Fig.\ref{fig:gaps_spins} we conclude that, for the practically interesting case
$\delta_0 \gnsim 0.15$, both methods give close positions of the critical point.

As was mentioned in Sec. \ref{sec:field_theory} in the presence of inplane rotational symmetry $b_1=b_2$ (e.g. frustrated Heisenberg model on the hexagonal lattice), quantum fluctuations become especially strong. 
In fact, when approaching the critical point $\rho_{cS}$ the integral $\int_q \frac{1}{\sqrt{\Delta^2 + K(q)}}\propto \int_q \frac{1}{\sqrt{\alpha^2 + (q^2-Q^2)^2}}$ is logarithmically divergent for $\alpha\rightarrow0$ at $q=Q$. It implies that one has to keep  higher order terms $\mathcal{O}(q_i^6)$ in the expansion (\ref{eq:K(q)})
\begin{equation}
K(\bm q) = \rho q^2 + \frac{b}{2}q^4 + c (q_x^6 + q_y^6) + d(q_x^4 q_y^2 + q_x^2 q_y^4)\label{eq:K6}
\end{equation}
which break the symmetry with respect to spatial rotations in the $\{xy\}$ plane and remove the degeneracy with respect to the choice of the direction of $\bm Q$. After accounting for the higher order anisotropic terms $\propto \mathcal O (q_i^6)$ the integral for $\langle \bm\pi^2 \rangle$ becomes  convergent at $|\bm q|=Q$ and the value $\rho_{cS}$ is well defined.

\section{Spin-spin correlation function}\label{sec:corr}

Spin-spin correlations of a standard tool to analyze quantum critical properties of a magnetic system.
In the SL phase  the correlator provides an essential information about the properties of the ground state.
The equal time two-point spin-spin correlation function reads 
\begin{eqnarray}\label{eq:C(r)}
C(r) = \langle n^\alpha(r) n^\alpha (0)\rangle = 1 + 2[R(r) - R(0)]
+\ldots,
\end{eqnarray}
where $\langle \pi^\alpha(r) \pi^\beta(0) \rangle = \delta^{\alpha\beta} R(r)$ and indices $\alpha,\beta$ refer only to the $x$ and $y$ spin components.
The two-point correlator is normalized such that $C(0)=\langle n_\alpha^2\rangle = 1$. In the 
SL phase the correlation function should vanish at large distances 
$C(r\rightarrow\infty)\rightarrow 0$ and $R(r\rightarrow\infty)\rightarrow 0$. These conditions are consistent with the ``melting criterion'' in Eq.(\ref{pc}) if we truncate the asymptotic expansion
in Eq.(\ref{eq:C(r)}) keeping only the terms explicitly presented there.

\begin{figure}
\includegraphics[scale=0.27]{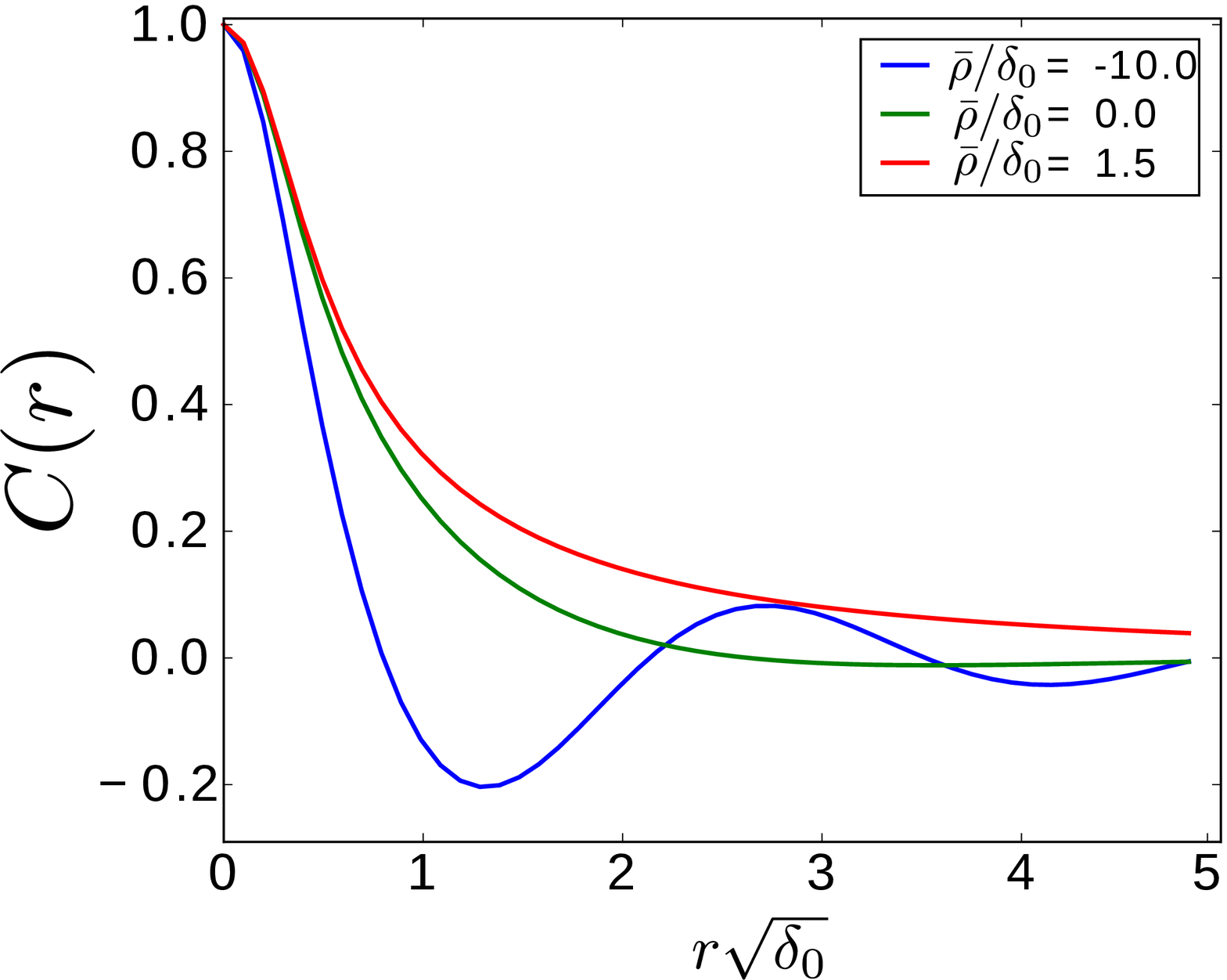}
\includegraphics[scale=0.28]{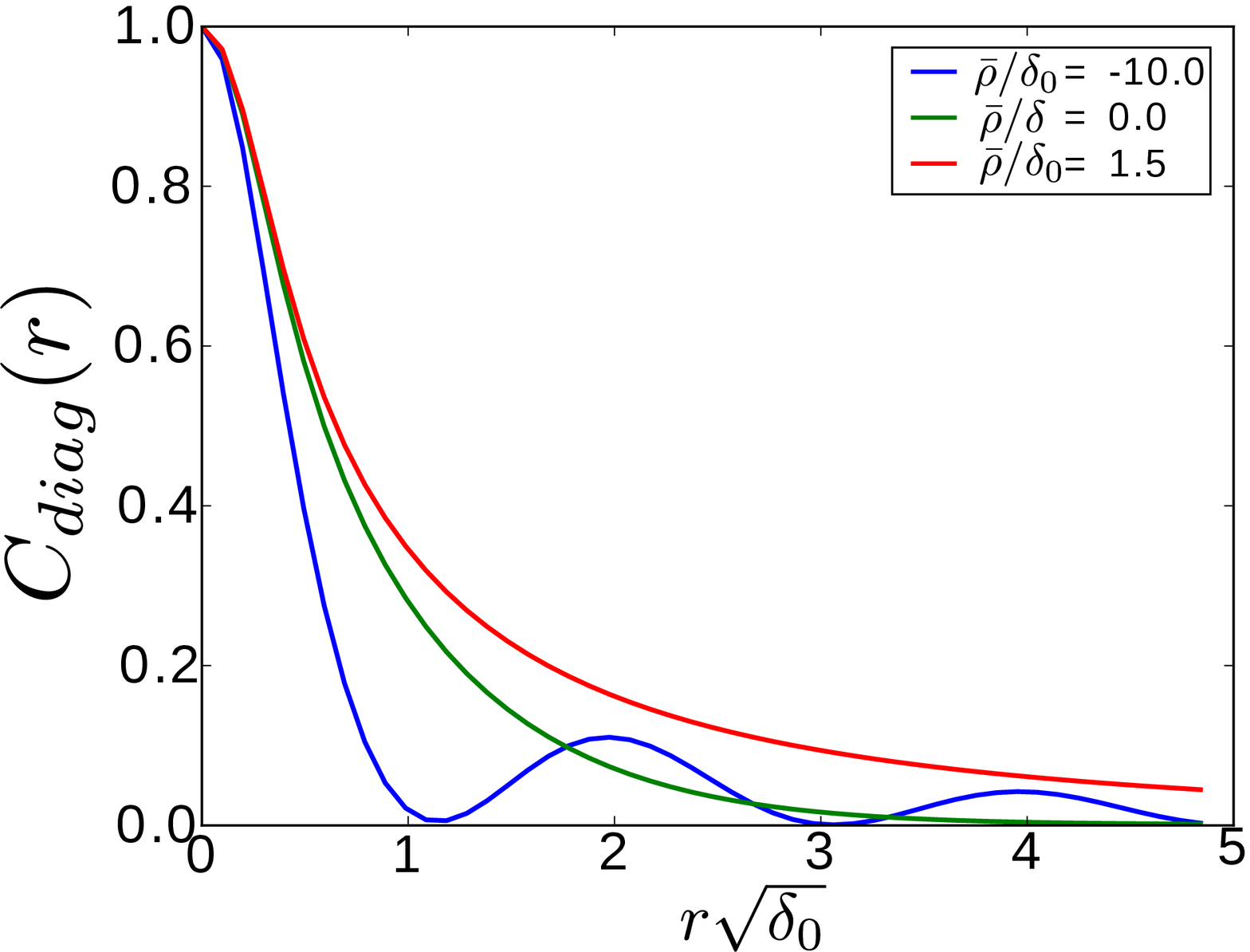}
\begin{picture}(0,0)
  \put(-125,225){ \text{a)}}
  \put(-125,95){\text{b)}}
\end{picture}
\caption{Static spin-spin correlation function $C(r)$ in the spin liquid phase for positive and negative spin stiffness ($b_2=0$, $\delta_0\approx0.04$). The radius vector $\bm r$ is directed along a) the principal lattice axes ($x$ or $y$), b) $\bm r$ is along the diagonal direction.}
\label{fig:Corr}
\end{figure}

The $\langle \bm \pi (r) \bm \pi (0) \rangle$ correlation function  in the SL phase reads
\begin{equation}\label{eq:R(r)}
R(r) = \int \frac{id\omega d^2q}{(2\pi)^3} \frac{e^{i \bm{qr}}}{\chi_\perp(\omega^2-\Delta^2) - K(\bm q)  + i0}.
\end{equation}  
Calculating (\ref{eq:R(r)}) and substituting the result in Eq. (\ref{eq:C(r)}), we obtain the 
two-point spin-spin correlation function $C(r)$; the numerical results are plotted in Fig (\ref{fig:Corr}).
Similar to the previous Section these plots correspond to the case $b_2=0$.
Therefore, the correlator is somewhat anisotropic.
There are two points to note, one is physical and another is technical.
 (i) The correlation length scales as one over the square root of the gap, $\xi \propto 1/\sqrt{\delta_0}$,
 instead of the standard relation, $\xi \propto 1/\delta_0$.
(ii) When integrating in Eq.(\ref{eq:R(r)}) we  use the soft ultraviolet cutoff by multiplying 
the integrand by $e^{-q^2/(2\Lambda^2)}$. The soft cutoff allows us to avoid nonphysical oscillations 
in $R(r)$ due to the Gibbs phenomenon. The Gibbs phenomenon results in spurious oscillations, which always exist for a sharp cutoff and are well known in
 Fourier analysis.

The asymptotic behaviour of the correlation function $R(r\rightarrow\infty)$ in the spin liquid phase at $\rho=0$ can be analytically obtained in the simplified isotropic approximation ($b_1=b_2$):
\begin{eqnarray}\label{eq:R_LP}
R(r) \sim \frac{e^{- r \sqrt{\frac{\delta_0}{2}}}}{r} \cos\left(r \sqrt{\frac{\delta_0}{2}}-\frac{\pi}{4}\right), 
\end{eqnarray}

Using Eq. (\ref{eq:R_LP}) we deduce  the spin-spin correlation length   $\xi=\sqrt{\frac{2}{\delta_0}}$. In the case of negative spin-stiffness ($\rho_{cS}<\rho<0$) the correlation function $R(r)$ becomes oscillating, see Fig. (\ref{fig:Corr}).
In the vicinity of the critical point $\rho_{cN}$ the correlations decay as 
\begin{equation}
R(r)=\frac{1}{2\pi \sqrt{2 \chi_\perp b_1}} I_0\left(r\frac{\sqrt{\bar\rho_{cN}}}{2}\right)K_0\left(r\frac{\sqrt{\bar\rho_{cN}}}{2}\right) \underset{r\rightarrow\infty}{\sim} \frac{1}{r}.\label{eq:R_cN}
\end{equation}
Formula (\ref{eq:R_cN}) is consistent with the well known $\propto 1/r$ decay  of  correlations of transverse spin components in the N\'eel phase (see e.g. Ref. \cite{Takahashi89}).
We stress that the ``isotropic approximation'', $b_1=b_2$, provides a qualitative and quantitative description of the correlation function $C(r)$ only away from the critical point $\rho_{cS}$. In the vicinity of the point $\rho_{cS}$ the isotropic model (\ref{eq:L}) becomes unstable, see comments to Eq. (\ref{eq:K6}).

Now we would like to make a comparison between $O(3)$ and $O(2)$ quantum Lifshitz transitions. The $O(2)$ version of Lagrangian (\ref{eq:L}) describes the XY frustrated Heisenberg antiferromagnet in the continuous limit.
The physics in the $O(2)$ model is quite different from the $O(3)$ model and the Ioffe-Larkin argument is inapplicable in this case.
 The $O(2)$ Lagrangian can be mapped to the scalar Lifshitz model described by a polar angle $\theta$: $n_x+in_y=e^{i\theta}$. This model has an exact solution for the correlation function $C(r)$ at the LP: $C(r)$ decays algebraically\cite{Ardonne04} at the LP in contrast to the non-vanishing correlations at $r\rightarrow\infty$ in long-range ordered N\'eel or spin-spiral phase. Therefore we conclude that there exist a finite region in the vicinity of the LP with algebraically decaying correlations. The region with algebraic spin correlations in some extent is analogous to the SL phase in the $O(3)$ model addressed in the present paper.

\section{$J_1-J_3$ model on the square lattice}\label{sec:J1-J3}
In the present Section we compare the field theory predictions with results of numerical calculations
for the  antiferromagnetic $J_1-J_3$ Heisenberg model on the square lattice.
Frustrated $J_1-J_2$ and $J_1-J_2-J_3$ models have been discussed in  numerous studies (see e.g. Refs. \cite{Ferrer93, Sindzingre10, Reuther11a}): some references are
also presented in the Introduction.
In the classical limit both models exhibit the spin spiral state at a sufficiently large frustration.
Quantum versions of the models show a magnetically disordered state at a sufficiently large frustration.
Classically the $J_1-J_2$ model at $J_2/J_1=1/2$ has three degenerate ground states,
the N\'eel, the spin-spiral, the spin-stripe. The tricritical point is somewhat special; the proximity of the columnar spin stripe phase enhances spin-dimer correlations and makes the physics of the $J_1-J_2$ model
different from that considered in the present work. On the other hand if we set $J_2=0$
and consider only the $J_3$ frustration then classically there is a Lifshitz
point with a transition to the spin-spiral at $J_3=J_1/4$, and  the spin-stripe state 
has much higher energy than the spin-spiral and the N\'eel states.
Therefore the $J_1-J_3$ model is a good testing ground
for the generic theory of a ``soft'' Lifshitz transition developed in the present work.
The Hamiltonian of the $J_1-J_3$ model reads
\begin{equation}
H = J_1 \sum_{<ij>} \bm S_i \bm S_j + J_3 \sum_{\langle\langle\langle ij \rangle\rangle\rangle} \bm S_i \bm S_j,
\end{equation}
where $<ij>$ and $\langle\langle\langle ij \rangle\rangle\rangle$ denotes first and third nearest neighbour interaction.
The classical spin-spiral to N\'eel LP is located at $J_3/J_1=1/4$.
As we already pointed out in Section \ref{sec:field_theory}
quantum fluctuations must shift the LP towards larger values $J_3/J_1>1/4$.

 In the long-wavelength approximation we can map the Heisenberg model to the Lagrangian (\ref{eq:L}).
The magnetic susceptibility is well known,
\begin{equation}
\label{cperp}
 \chi_{\perp}=\frac{1}{8J_1}.
\end{equation}
The elasticity parameters  of the Lagrangian can be found in two ways.
(i) The first way is a straightforward expansion of the classical elastic energy
at small wave number $q$, that gives
\begin{eqnarray}
\label{rb}
&&\rho=S^2(J_1-4J_3),\nonumber\\
&&b_1=S^2\frac{(16J_3-J_1)}{12},\nonumber\\
&&b_2=0.
\end{eqnarray}

(ii) An alternative way is to calculate the magnon dispersion in the N\'eel phase
using the standard spin-wave theory. The dispersion reads~\cite{Ferrer93}:
\begin{eqnarray}\label{eq:spin-wave}
&&\omega_q = 4S J_1 \sqrt{\left(1-\frac{J_3}{J_1}(1-\gamma_{2q})\right)^2-\gamma_q^2},\\
&&\gamma_{q} = \frac{1}{2}(\cos{q_x}+\cos{q_y}),\nonumber\\
&&\gamma_{2q} = \frac{1}{2}(\cos{2q_x}+\cos{2q_y}).
\end{eqnarray}
Expanding $\omega_q$ at small $q$ and comparing the results with Eq.(\ref{eq:omega_q}) (at $\Delta=0$)
we find
\begin{eqnarray}
\label{rb1}
&&\rho=S^2(J_1-4J_3),\nonumber\\
&&b_1=4 J_1 S^2\left[-\frac{5}{48}+\frac{2}{3}\left(\frac{J_3}{J_1}\right)+\left(\frac{J_3}{J_1}\right)^2\right],\nonumber\\
&&b_2=4 J_1 S^2\left[-\frac{1}{8}+2\left(\frac{J_3}{J_1}\right)^2\right].
\end{eqnarray}

Expressions for $b_1$ and $b_2$ in Eqs.(\ref{rb}) and (\ref{rb1}) do not coincide.
At the LP, $J_3=J_1/4$, both Eqs. give $b_2=0$,  however, values of $b_1$ are different,
Eq.(\ref{rb}) gives $b_1=0.25S^2J_1$ while Eq.(\ref{rb1}) gives $b_1=0.5S^2J_1$.
Of course the spin-wave theory value is more reliable.

\begin{figure}
\includegraphics[scale=0.2]{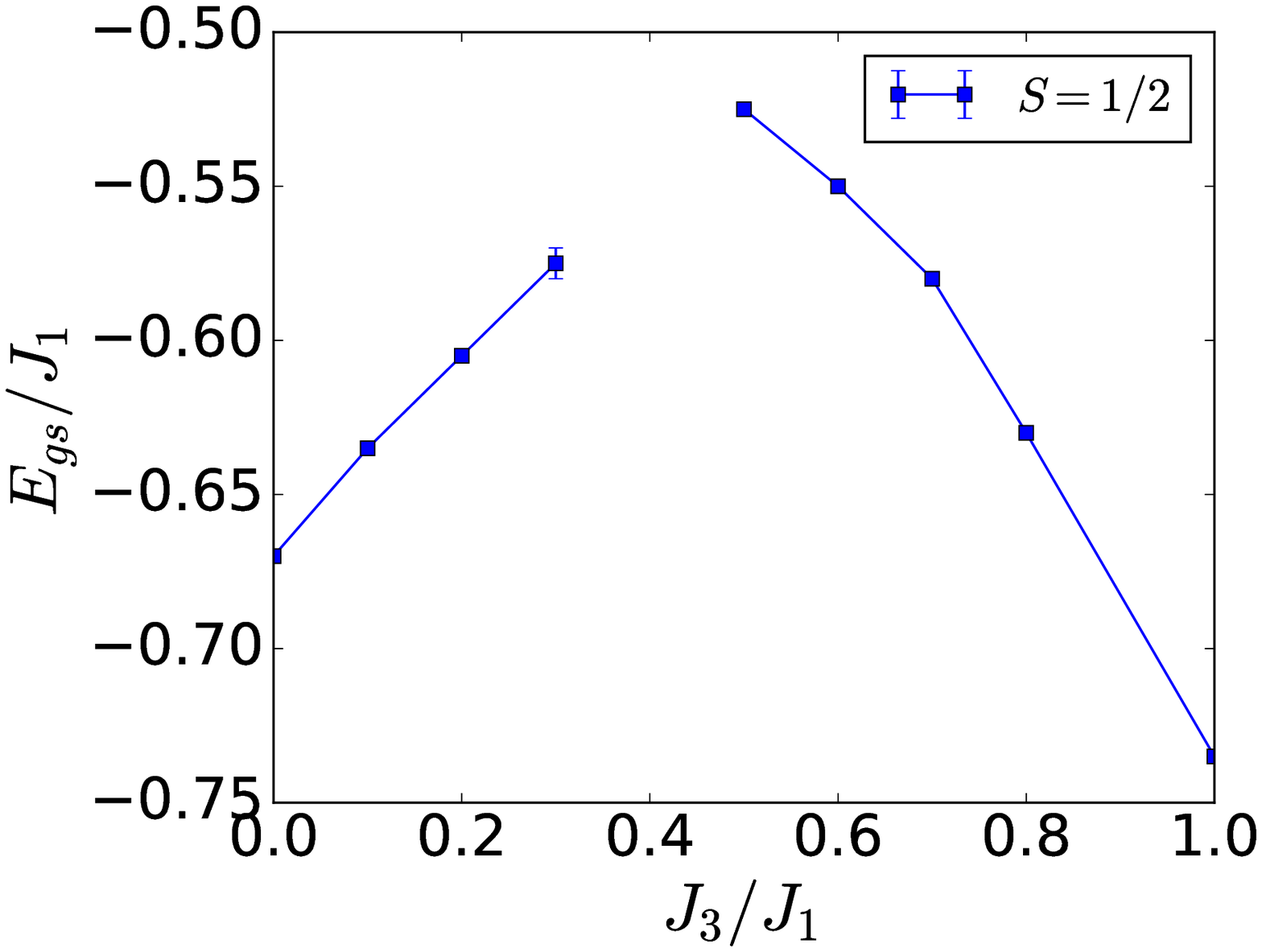}
\includegraphics[scale=0.2]{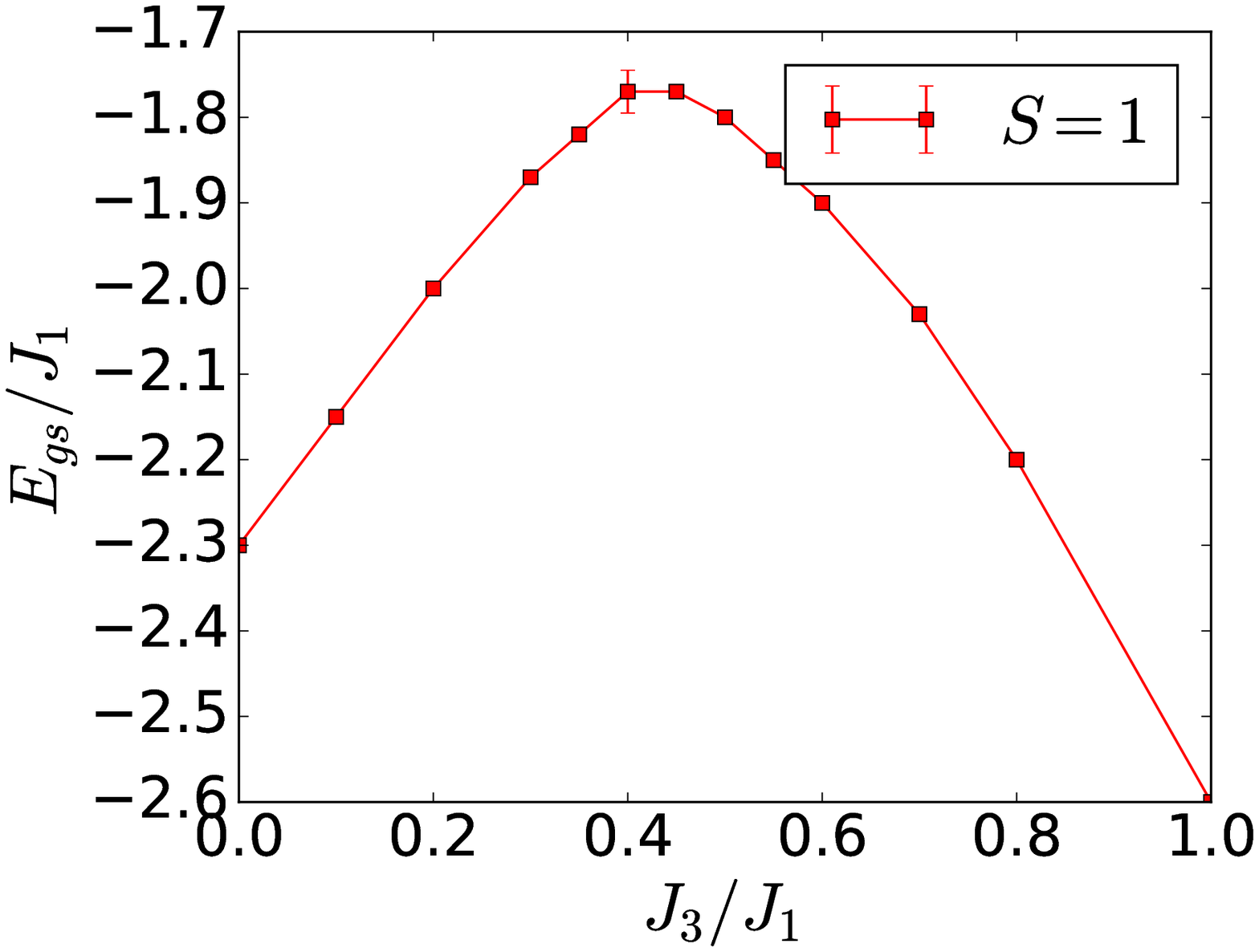}
\begin{picture}(0,0)
  \put(-211,67){ \text{a)}}
  \put(-93,67){\text{b)}}
\end{picture}
\caption{ $J_1-J_3$ model ground state energy in the N\'eel and in the Spin Spiral states for a) $S=1/2$ and b) $S=1$ calculated by numerical series expansion method.
}
\label{gse}
\end{figure}

\begin{figure}
\includegraphics[scale=0.3]{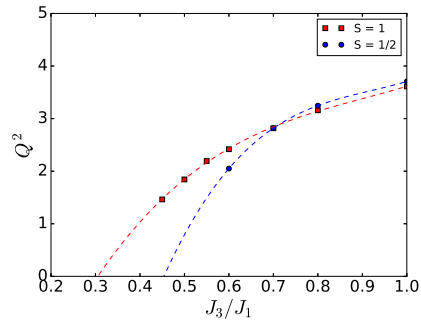}
\caption{ Spiral wave vector (squared) $Q^2$ versus $J_3$. Dots show results of numerical series 
expansion. Blue (red) dots correspond $S=1/2$ ( $S=1$).
Dashed lines show fits of data by cubic polynomials, $Q^2=a_1(J_3-J_3^{LP})+a_2(J_3-J_3^{LP})^2 + a_3(J_3-J_3^{LP})^3$.
}
\label{fig:Q_rho}
\end{figure}

We have performed extensive series calculations both in the N\'eel phase  and the spin-spiral phase.
Unfortunately the series expansion method does not allow to assess properties of the 
spin liquid phase directly.
However, it allows to estimate the range of parameters where the spin liquid exists which can be
compared with predictions of the field theory.
In the N\'eel phase the series starts from the simple Ising antiferomagnetic state.
In the spiral phase the calculation is more tricky. We first impose a classical diagonal spiral with
some  wave vector $Q$ and find the total energy of this state $E(Q)$. This includes the classical 
energy and the
quantum corrections calculated by means of series expansions. We perform this calculations for many
values of $Q$ and then find numerically the minimum of $E(Q)$. Such procedure gives us the
ground state energy $E_{gs}$ and the physical wave vector $Q$.
The ground state energy $E_{gs}$ is plotted in Fig. \ref{gse}  versus $J_3$.
The plot of the wave vector squared, $Q^2$, versus $J_3$ is presented in Fig. \ref{fig:Q_rho}.
From the field theory description we expect that near the LP the wave vector behaves as
\begin{eqnarray}
\label{QQa}
Q^2=\frac{2|\rho|}{b_1}=\frac{8S^2}{b_1} ({J_3}-J_3^{LP}).
\end{eqnarray}
Therefore, from Fig. \ref{fig:Q_rho} we determine positions of Lifshiz points and, using Eq.(\ref{QQa}) we find the values of the elastic constant $b_1$
at the LP:
\begin{eqnarray}
\label{LP}
&& S=1/2: \ \ \ J_3^{LP} \approx 0.45 J_1, \ \ \ b_1/S^2\approx 0.60 J_1,   \nonumber\\
&& S=1: \ \ \ \ \ \ J_3^{LP} \approx 0.3 J_1,  \ \ \ b_1/S^2\approx 0.74J_1.
\end{eqnarray}

As expected, (see the very end of Section \ref{sec:field_theory}),
quantum fluctuations extend the N\'eel phase in relation to the classical LP 
$J_3^{LP}=0.25J_1$. Values of the elastic constant $b_1$ are larger than that given
by Eq.(\ref{rb}) and smaller than that given by Eq.(\ref{rb1}).

\begin{figure}
\includegraphics[scale=0.33]{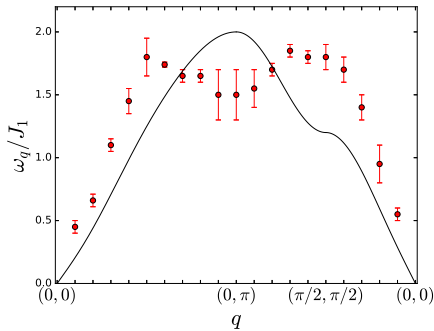}
\caption{ Magnon dispersion $\omega_q$  for $J_1-J_3$ model on the square lattice in 
the N\'eel phase at $J_3/J_1=0.2$. Red circles correspond to the series expansion results, 
black line is the  linear spin-wave dispersion in Eq. (\ref{eq:spin-wave}). }
\label{fig:omega_q}
\end{figure}

We have also calculated the magnon dispersion in the N\'eel phase.
The series expansion becomes erratic at $J_3 > 0.2J_1$ and the errorbars in the calculations of  $\omega_q$ grow very quickly.
The dispersion at $J_3 = 0.2J_1$ is shown in Fig. \ref{fig:omega_q}. 
We see that the shape of the dispersion is somewhat different from the prediction of the
spin-wave theory (\ref{eq:spin-wave}). On the the other hand the total bandwidth
is consistent with the spin-wave theory. The situation is different in the case of a simple
Heisenberg model ($J_3=0$), when the shape of magnon dispersion is consistent with the spin-wave theory
but the total bandwidth is about 20\% larger compared to the spin-wave theory value.

\begin{figure}
\includegraphics[scale=0.39]{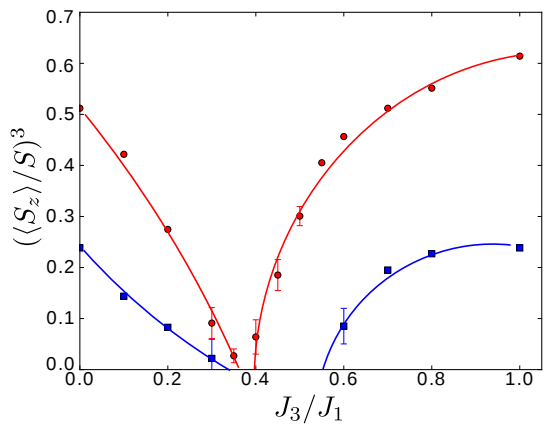}
\caption{ Average onsite magnetization cubed. Blue squares (red circles) show 
series expansion results for $S=1/2$ ($S=1$), 
solid lines are guides for the eye. }
\label{fig:M3}
\end{figure}

We also compute the static on-site magnetization in the N\'eel and spiral phases.
The magnetization vanishes at $J_3^{cN}$ and $J_3^{cS}$ critical points.
We already pointed out that the N\'eel-SL transition at $J_3^{cN}$ belongs to the $O(3)$ universality class.
Therefore, we expect
scaling
$\langle S_z\rangle \propto |J_3-J_3^{cN}|^\beta $ when approaching  the critical point from the N\'eel phase, here 
$\beta = (D-2+\eta)\nu/2 \approx \nu/2 \approx 0.35$.
Due to this reason in  Fig. \ref{fig:M3} we show series expansion results for the 
static on-site magnetization cubed.
From here we locate the critical points.
\begin{eqnarray}
\label{cp1}
&& S=1/2: \ \ \ J_3^{cN} \approx 0.35 J_1, \ \ \ J_3^{cS}\approx 0.55J_1,   \nonumber\\
&& S=1: \ \ \ \ \ \ J_3^{cN} \approx J_3^{cS} \approx 0.35J_1.
\end{eqnarray}

Our result for the SL range $\Delta J_3$ in the case $S=1/2$  is different from the  recent 
work \cite{Reuther11a}, that suggest the SL phase at 
$0.4 \leq J_3/J_1 \leq 0.8$. However, our predictions are reasonably close to the exact diagonalization results \cite{Sindzingre10}, suggesting the gapped SL phase for $0.45 \leq J_3/J_1 \leq 0.65$.
Note also that the critical index for the the $ J_3^{cS}$ critical point
is smaller than the $O(3)$ value, $M \propto (J_3-J_3^{cS})^{\beta}$, $\beta \sim 0.2$. 

Now we can compare the results of series calculations with predictions
of the field theory. Eqs.(\ref{cperp}),(\ref{LP}) give values of
$\chi_{\perp}$ and $b_1$. Hence, according to Eqs.(\ref{eq:delta_0}) and (\ref{rd})
values of the gap at  the LP are
\begin{eqnarray}
\label{rb2}
&& S=1/2: \ \ \ \ \delta_0 \approx 0.66, \quad \Delta_0 \approx 0.53 J_1,\nonumber\\
&&S=1: \ \ \ \ \ \ \ \delta_0 \approx 0.17, \quad \Delta_0 \approx 0.29 J_1 \ .
\end{eqnarray}

Formally the field-theoretical prediction (\ref{rd}) is derived within logarithmic accuracy and valid at $\delta_0  \ll 1$, while these values, especially that
at $S=1/2$, are not small. Nevertheless, we believe that Eq.(\ref{rb2}) gives a reasonable
estimate of the gaps. 
Knowing the dimensionless gaps and using Fig. \ref{fig:gaps_spins} we can deduce the window
$\delta{\bar \rho}$ occupied by the spin liquid phase. Combining this with Eq.(\ref{QQa}) we find
the spin liquid window $\Delta J_3=|J_3^{cS}-J_3^{cN}|$ that follows from the field theory,
\begin{eqnarray}
 &&\Delta J_{3}/J_1  \approx 0.3, \quad (S=1/2),\nonumber \\
 &&\Delta J_{3}/J_1\approx 0.1, \quad (S=1) \ . 
\label{eq:deltaJ3}
\end{eqnarray}
These values while being slightly larger are in a reasonable agreement with the SL phase windows following from series expansion data in
Fig. \ref{fig:M3}.

In conclusion of this Section we would like to comment on the anisotropic $J_1-J_3$ model
on square lattice. \cite{Oitmaa16}
In this model $J_3$ frustrates $J_1$ only in one direction, say $J_3$ connects only the
third nearest neighbours in the $y$-direction.
This results in an  anisotropic LP: the spin stiffness $\rho_y$ vanishes
at some value of $J_3$ while $\rho_x$ remains finite and positive.
The wave vector of the spin spiral is always directed along the y-axis.
In this case quantum fluctuations at the LP are described as
$\langle \bm\pi^2\rangle \propto \int \frac{d^2 q}{ \sqrt{q_x^4 + q_y^4 + \rho_x q_x^2} }$.
The integral is infrared convergent unlike that in the isotropic LP.
Therefore generically one cannot expect a spin liquid in this case.
The fluctuations are still enhanced and there must be a suppression
of the on-site magnetization at the LP.  This is exactly what series expansions for
the anisotropic $J_1-J_3$ model with S=1/2 indicate. \cite{Oitmaa16}
It is likely that a similar scenario is valid for thin films of frustrated manganites 
(Tb,La,Dy)MnO$_3$ tuned close to LP.

\section{Conclusion}\label{sec:concl}
In this work, using field theory techniques, we have studied properties of the universal spin 
liquid phase in a vicinity of an isotropic Lifshitz point in a system of localized frustrated spins.
Our general analysis includes the phase diagram, positions of critical points, excitation spectra, 
and spin-spin correlations functions.
In the semiclassical regime of large spin S the spin liquid phase forms an exponentially narrow region in the vicinity of the Lifshitz point.
The derivation of these results is accompanied with a thorough discussion of the criterion for quantum melting of long range
magnetic order in two dimensions, an analogue of Lindemann criterion.
We argue  the 2D Lifshitz point spin liquid is similar to the gapped Haldane phase in integer-spin 
1D chains.
In order to check our general field theory results, and in particular to check the quantum melting criterion,
we have performed numerical series expansion calculations for the $J_1-J_3$ model on square lattice.
We demonstrate that results of these two different approaches are in a good agreement.

\section{Acknowledgments}
We would like to thank G. Khaliullin for insightful comments and suggestions.
The work has been supported by Australian Research Council No DP160103630.

\newpage
\appendix

\section{The value of $\langle \pi^2\rangle_c$ derived from
asymptotic Taylor expansion.}\label{sec:nz_Taylor}

After expanding $n_z=\sqrt{1-\bm\pi^2}$ in a Taylor series and using Wick's theorem:
\begin{eqnarray}\label{eq:nz_Taylor}
\langle n_z\rangle = 1 - \sum_{k=1}^{\infty} \langle \bm\pi^2 \rangle^k \frac{ (2k-2)!}{2^{2k-1} (k-1)!} \nonumber \\ =
1 - \frac{1}{2}\langle \bm\pi^2\rangle - \frac{1}{4}\langle \bm\pi^2\rangle^2 - \frac{3}{8}\langle \bm\pi^2\rangle^3 + \ldots.
\end{eqnarray}

The series (\ref{eq:nz_Taylor}) is asymptotic and the coefficients at large $k$ diverge.  Since the series is asymptotic we truncate it when the coefficients in front of $\langle\bm\pi^2\rangle^k$ terms become larger then unity. Accounting for the leading terms in the expansion up to $\langle \bm \pi^2\rangle^3$ inclusive gives the critical value $\langle \bm\pi^2\rangle_c \approx 0.93$ for $\langle n_z\rangle=0$.

\section{Excitations in static spin-spiral phase}\label{sec:append_spiral}
By considering fluctuations in the spin spiral state we find
the condition when quantum fluctuations melt the spiral.
Here we derive the dispersions of in plane and out of plane fluctuations in the spin-spiral state. 
To be specific let us assume that the spiral lies in $\{xy\}$ plane:
\begin{eqnarray}
\label{spi}
{\bm n}=(\cos{\bm Q}{\bm r},\sin {\bm Q}{\bm r},0) \ .
\end{eqnarray}
There are two different spin waves, the in-plane $\varphi({\bm r},t)$,\begin{eqnarray}
\label{wfi}
{\bm n}=(\cos({\bm Q}{\bm r}+\phi),\sin({\bm Q}{\bm r}+\phi),0) \ ,
\end{eqnarray}
and the out-of-plane $h({\bm r},t)$,
\begin{eqnarray}
\label{wh}
{\bm n}=(\sqrt{1-h^2}\cos{\bm Q}{\bm r},\sqrt{1-h^2}\sin{\bm Q}{\bm r},h) \ .
\end{eqnarray}
Substituting parametrization (\ref{wfi}) and (\ref{wh}) in the Euler-Lagrange equations of motion corresponding to the Lagrangian (\ref{eq:L}) and linearising the equations with respect to $\phi$ and $h$ we obtain the dispersion of the in-plane and out of plane modes. The derivation is straightforward, see e.g. Ref. \cite{Milstein15}. The dispersion of the in-plane mode is
\begin{eqnarray}
\label{dfi}
\omega_{\bm q}^2 = \frac{1}{\chi_\perp}\left[ K(\bm Q) - \frac{1}{2}\left(K(\bm Q + \bm q) + K(\bm Q - \bm q)\right)\right] \nonumber\\
=\frac{b_1}{2\chi_{\perp}}\left[2Q^2q^2+q_x^4+q_y^4\right] \ ,
\end{eqnarray}
and the dispersion of the out-of-plane mode is
\begin{eqnarray}
\label{dh}
\Omega_{\bm q}^2 = \frac{1}{\chi_\perp}\left[K(\bm q) - K(\bm Q)\right] \nonumber\\ 
=\frac{b_1}{2\chi_{\perp}}\left[Q^4/2-Q^2q^2+q_x^4+q_y^4\right] \ .
\end{eqnarray}
The total quantum fluctuation orthogonal to the spin alignment in the spiral phase reads
\begin{eqnarray}
\label{qfb}
&&\langle \bm\pi^2 \rangle= \langle \phi^2 \rangle+\langle h^2 \rangle, \\
&&\langle \phi^2 \rangle=
\int\frac{d^2q}{(2\pi)^2}\frac{1}{2 \omega_{\bm q}}, \nonumber\\
&&\langle h^2 \rangle=
\int\frac{d^2q}{(2\pi)^2}\frac{1}{2 \Omega_{\bm q}}. \nonumber
\end{eqnarray}
From the condition $\langle \bm\pi^2 \rangle = \langle \bm \pi^2 \rangle_c \approx 1$ we find the position of the spiral-SL critical point $\rho_{cS}$, see Sec. \ref{sec:crit_rho_delta} in the main text.

%
%
%
%
%
%
%
%
%
%
%
%
%
%
%
%
%

\end{document}